\documentclass{article}

\begin{document}

\newcommand{\dydx}[2]{\frac{\partial{#1}}{\partial{#2}}}
\newcommand{\dydxt}[2]{\frac{d{#1}}{d{#2}}}
\newcommand{\DD}{\Delta}
\newcommand{\GG}{\Gamma}

\newcommand{\ppp}{\partial}

\newcommand{\be}{\begin{eqnarray}}
\newcommand{\ee}{\end{eqnarray}}
\newcommand{\bes}{\begin{eqnarray*}}
\newcommand{\ees}{\end{eqnarray*}}

\title{Causality and Peierls Bracket in Classical Mechanics}
\author{Pankaj Sharan\thanks{email : pankaj.ph@jmi.ac.in}\\
Physics Department, Jamia Millia Islamia,\\ New Delhi 110 025, INDIA}

\maketitle
\begin{abstract} 
Relation between the Peierls and the Poisson bracket is derived in classical 
mechanics of time-dependent systems.
Equal-time Peierls brackets are seen to be the same 
as the Poisson brackets in simple cases but a proof for a general Hamiltonian is lacking.
\end{abstract}

\section{Introduction}
Once the coordinates and momenta of an autonomous system are specified at any fixed time, 
the phase trajectory of the system is determined for all past and future. But for a system
in the presence of external agents, the phase trajectories get affected by
these external agents only {\em after} the time they are switched on. This requires
a formulation of the condition of causality in classical mechanics although it is always
assumed implicitly. In this note we 
find that a bracket, defined by Peierls\cite{peierls} in 1952 as a covariant
bracket for relativistic fields, is the natural bracket 
for implementing causality in general time-dependent classical systems.

In the autonomous case observables are functions of coordinates and momenta. The time dependence
of such an observable is given by the values of the observable along a phase trajectory 
which is determined by Hamiltonian equations of motion. In the case of time-dependent systems the definition
of an observable has to be extended to include time as an extra variable on 
which it can depend. This necessitates the inclusion of time as a dynamical 
variable alongwith coordinates and momenta.

The Poisson bracket determines how one quantity $b(t,q,p)$ changes 
another quantity $a(t,q,p)$ when it acts as the 
Hamiltonian or vice-versa. The Peierls bracket, on the other hand, 
determines how one quantity $b(t,q,p)$ when
{\em added} to the system Hamiltonian $h$ with an infinitesimal coefficient $\lambda$
affects changes in another quantity $a(t,q,p)$ and vice-versa. Therefore, we expect
a close relationship between the two brackets when the system Hamiltonian is zero.
This indeed is true as we prove in section \ref{proof} below. We find that the Peierls bracket
of any two observables for zero Hamiltonian is related to `two-point' Poisson bracket 
of time dependent quantities and reduces to the Poisson bracket at equal times. 

What is even more interesting is that even for non-zero Hamiltonians the canonical Peierls brackets 
at equal-times coincide with the canonical Poisson brackets. For {\em unequal} times the Peierls
bracket is different from the two-point Poisson brackets and is characteristic of the governing
Hamiltonian.  
We can see this for simple Hamiltonians like free particle
and Harmonic oscillator or quadratic Hamiltonians. But a proof (or a counter example) of the fact that
equal-time Peierls bracket is the same as Poisson bracket for a general Hamiltonian
seems to be lacking.

\section{Phase trajectories and observables}
For simplicity of notation we consider just one coordinate and its momentum. The 
case of $N$ degrees of freedom is similar.

The phase trajectories are curves $\sigma:  t\to (t,q=F(t),p=G(t)$ from one dimensional
time manifold $T$ into the 
extended phase space $\GG$ with coordinates $(t,q,p)$.
The possible trajectories $\sigma$ are solutions to
\be \dydxt{F}{t}=\dydx{h}{p},\qquad
\dydxt{G}{t}=-\dydx{h}{q}, \ee
where $h(t,q,p)$ is the Hamiltonian function.

The rate of change of an observable $a(t,q,p)$ {\em along a phase trajectory} $\sigma$ 
is determined by 
\bes \left.\dydxt{a}{t}\right|_\sigma=
\left.\left(\dydx{a}{t} + \{a,h\}\right)\right|_{q=F(t),p=G(t)}.\ees
The Poisson bracket $\{a,b\}$ of two observables $a(t,q,p)$ and $b(t,q,p)$ is defined as 
\bes \{a,b\} = \left(\dydx{a}{q}\dydx{b}{p}-\dydx{a}{p}\dydx{b}{q}\right).\ees
The Poisson bracket as a function of $t,q,p$ refers to a single time t.  
Its explicit time dependence, if any, comes from that of $a$ and $b$.

For time-dependent systems it is useful to define the {\em value} of an observable
$a(t,q,p)$ on a phase trajectory as the integral $\int a(t,q,p)dt$. We assume that an observable
is always multiplied by (`smeared with')  an appropriate switching function of 
$t$ for the integral to exist. If $a(t,q,p)$ has
a Dirac delta function $\delta(t-t_a)$ as a factor, 
the value of $a$ of a will reduce to the value of
the observable at a specific time $t_a$ on the trajectory.

\section{\label{proof} Peierls bracket}

The Peierls bracket was originally introduced for relativistic fields
by Peierls\cite{peierls} in 1952 and has been promoted extensively by B. S. DeWitt\cite{dewitt}
as the fundamental covariant object in quantum field theory.

While the Poisson bracket between two observables $a$ and $b$
is defined on the whole phase space and is not dependent on the
existence of a Hamiltonian, the Peierls 
bracket refers to a specific trajectory determined by a governing Hamiltonian. The
Peierls bracket is related to
the change in an observable when the trajectory on which it is evaluated
gets shifted due to an infinitesimal change in
the Hamiltonian of the system by another observable. 

Let the Hamiltonian $h$ be deformed by a 
term $\lambda b$ where $\lambda$ is a small parameter and $b(t,q,p)$ an observable. Due to   
this term the phase trajectory gets modified by amounts proportional to $\lambda$.

We can compare the values $\int adt$ of an observable $a$ on the two trajectories $\sigma_0$ and  
$\sigma_{\lambda b}$ :
\bes \sigma_0 : t\to q &=& F_0(t), p=G_0(t),\\  
\sigma_{\lambda b} : t\to q &=& F(t)=F_0(t)+\lambda F_b(t),\\
 p &=& G(t)=G_0(t)+\lambda G_b(t),\\
\lim_{t\to -\infty}F(t) &=& F_0(t),\qquad \lim_{t\to -\infty}G(t) = G_0(t)
\ees
which agree in remote past and then evolve with the Hamiltonians $h$ and $h+\lambda b$
respectively. The change in $\int a(t,q,p)dt$ as $\lambda\to 0$ is written
\be D_ba \equiv \lim_{\lambda\to 0}
\frac{1}{\lambda}\left[\int_{\sigma_{\lambda b}}adt-\int_{\sigma_0}adt\right].\ee  
Similarly, we can find the change in $b$ when $a$ deforms the Hamiltonian.
The Peierls bracket is given by 
\be [a,b]\equiv D_ba-D_ab\ee
For the Hamiltonian $h+\lambda b$ the phase trajectories are 
\bes \dydxt{F}{t} &=& \left.\left(\dydx{h}{p}+\lambda\dydx{b}{p}\right)\right|_{q=F(t),p=G(t)}\\
\dydxt{G}{t} &=& -\left.\left(\dydx{h}{q}+\lambda\dydx{b}{q}\right)\right|_{q=F(t),p=G(t)}\ees
There are  similar equations for $F_0,G_0$ when $\lambda=0$.

Now, neglecting $O(\lambda^2)$, 
\bes \left.\left(\dydx{h}{p}+\lambda\dydx{b}{p}\right)\right|_{F,G}
= \left.\left(\dydx{h}{p}+\lambda\dydx{b}{p}\right)\right|_{F_0,G_0}
+\lambda\left.\left(\dydx{^2h}{q\ppp p}F_b+\dydx{^2h}{p^2}G_b\right)\right|_{F_0,G_0}.
\ees
There is a similar  expression for $dG/dt$. Thus, 
\be \label{FBeqn} 
\dydxt{F_b}{t} &=& \left(\dydx{^2h}{q\ppp p}F_b+\dydx{^2h}{p^2}G_b\right)+\dydx{b}{p}, \nonumber\\
\dydxt{G_b}{t}&=&-\left(\dydx{^2h}{q^2}F_b+\dydx{^2h}{p\ppp q}G_b\right)-\dydx{b}{q}.\ee
The Peierls bracket will depend in general on the governing Hamiltonian $h$.
To relate to the Poisson bracket, where $b$ and $a$
act as Hamiltonians to each other, consider the case $h=0$.
Then the original trajectories are $q=F_0$=constant, $p=G_0$=constant and we can write 
\bes F_b(t)&=&\int_{-\infty}^t\dydx{b}{p}(t')dt\\
&=&\int\theta(t-t')\dydx{b}{p}(t')dt,\\
G_b(t)&=&-\int_{-\infty}^t\dydx{b}{p}(t')dt\\
&=&-\int\theta(t-t')\dydx{b}{q}(t')dt,\ees
where causality is automatically taken into account by
the step function ($\theta(t-t')=1$ for $t>t'$ and $\theta(t-t')=0$ for $t<t'$).
The values of $q$ and $p$ are fixed by the original trajectory $q=F_0,p=G_0$. 
The change in an observable $a$ is
\bes D_ba &=& \int \left(\dydx{a}{q}(t)F_b(t)+\dydx{a}{p}(t)G_b(t)\right)\\
&=& \int_{-\infty}^{\infty}dt\int_{-\infty}^{\infty} dt'\,\theta(t-t')\times\\
&& \left[\dydx{a}{q}(t)\dydx{b}{p}(t')-\dydx{a}{p}(t)\dydx{b}{q}(t')\right].\ees
The expression for $D_ab$ is similar with the roles of $a$ and $b$ interchanged.
The two step functions add to unity ($\theta(t-t')+\theta(t'-t)=1$)
on calculating $D_ba-D_ab$. The Peierls
bracket for the value of observables $a$ and $b$ is thus seen to be an integrated
`two-point' Poisson bracket with the time dependence at the individual times of the
observables : 
\be [a,b] = \int\int dt\,dt'\,\{a(t),b(t')\}_{\sigma_0} \ee
where
\bes \{a(t),b(t')\}_{\sigma_0}=\dydx{a}{q}(t)\dydx{b}{p}(t')-\dydx{a}{p}(t)\dydx{b}{q}(t')\ees
evaluated at the original trajectory $\sigma_0$ with constant $q=F_0,p=G_0$.

If the observables  $a$ and $b$ are localized in time, 
\bes a(t,q,p)=\delta(t-t_a)\alpha(t,q,p),\ b(t,q,p)=\delta(t-t_b)\beta(t,q,p), \ees
then the expression for the Peierls bracket is the same as Poisson bracket evaluated at
$q=F_0,p=G_0$ :
\bes [a,b]=\{\alpha(t_a,q,p),\beta(t_b,q,p)\}. \ees
For equal times it is just the Poisson bracket.
In general (when $h$ is not zero) the Peierls bracket will depend on the second derivatives of
the Hamiltonian as seen by the equation \ref{FBeqn} above for $F_b,G_b$. In particular, the $[q(t),q(t')]$ 
Peierls bracket will not be zero for different times and would be equal to the Poisson bracket only
at equal time. Simple examples in the next section illustrate this.

\section{Examples}

As one example take $h=(1/2)(q^2+p^2)$. 

Let  $b=q j$ where $j : t\to j(t)$ is a function
`switching on' the observable $q$ in some interval.
If $t\to q=F(t),p=G(t)$ is the deformed trajectory, then from equation (\ref{FBeqn}) 
\bes \dot{F}_b=G_b,\qquad \dot{G}_b=-F_b-j, \ees  
and the solution for the deformation $F_b$ is 
\bes F_b(t)=-\int G_R(t-s) j(s) ds \ees
where $G_R$ is the 
retarded Green's function satisfying 
\bes \ddot{G}_R(t)+G_R(t)=\delta(t).\ees
$G_R(t)$ is equal to $\sin t$ for $t>0$ and is $0$ for $t<0$.
Therefore, 
\bes F_b(t) &=& -\int G_R(t-s) j(s) ds, \\
G_b(t) &=& -\int \ppp_t G_R(t-s) j(s) ds. \ees

If $a$ is the other observable $a=q k$ (with the switching function $k$) then
\bes   D_ba=-\int\int k(t)G_R(t-s) j(s) dsdt. \ees
The Peierls bracket of $a$ and $b$ is  
\bes [a,b] &=& -\int\int[k(t)G_R(t-s) j(s)\\
&& +j(t)G_R(t-s)k(s)]dsdt\\
&=& -\int\int k(t)[G_R(t-s)-G_A(t-s)] j(s)\\
&=& -\int\int k(t)\sin(t-s)j(s),\ees
where $G_A(t)=G_R(-t)$ is the corresponding advanced Green's function.
If $k$ and $j$ were both Dirac delta functions supported at $t=t_a$ and $s=t_b$ 
then this Peierls bracket gives 
\bes [a,b] = -\sin(t_a-t_b),\ees
which is the two-point bracket of $q$ with $q$, depending on phase space points
of the original trajectory at times $t_a$ and $t_b$. It is zero
at equal times. Physically, since both $a$ and $b$ are
coordinates, the Peierls bracket measures how much of canonical momentum is
generated in $q$ from time $t_a$ to $t_b$ along the oscillator trajectory
if $t_b$ is later than $t_a$ or vice versa in the opposite case.

If $c=lp$ is another observable with $l(t)$ as the switching function 
then
\bes G_c(t) &=& -\int G_R(t-s) l(s) ds, \\
F_c(t) &=& -\dot{G_c}(t)=\int \ppp_tG_R(t-s) l(s) ds. \ees
Thus the `canonical' Peierls bracket between $a=qk$ and $c=pl$ is
\bes [a,c]=\int\int k(t)\cos(t-s) l(s)dsdt \ees
which gives $[q,p]=1$ at equal times when $a$ and $c$ are localized in time.
The Peierls bracket in this case measures how much of canonical momentum is
retained in $p$ from time $t_a$ to $t_c$ along the oscillator trajectory
if $t_c$ is later than $t_a$ or how much `coordinateness' is
retained in $q$ from time $t_c$ to $t_a$ if opposite is the case. 

We can check that the equal time Peierls brackets remain unchanged and equal to Poisson
brackets for some other simple Hamiltonians. For a free particle $h=p^2/2$,
the trajectories are $G_0=$ constant, $F_0(t)=G_0 t+F_1$ where $F_1$ is another constant. 
Let $b=qj$ as before. Then $\dot{F}_b=G_b$ and $\dot{G}_b=-j$. The solution is
\bes F_b(t) &=& -\int (t-t')\theta(t-t')j(t')dt'\\
G_b(t) &=& -\int \theta(t-t')j(t')dt'
\ees
which gives 
\bes [a,b]=-\int\int k(t)(t-t')j(t')dt'\ees
for $a=qk$, and 
\bes [c,b]=-\int\int l(t)j(t')dt'\ees
for $c=pl$. Again the equal-time canonical Peierls bracket is the same as the Poisson bracket.

It would be interesting
to prove that the equal time canonical Peierls bracket is the same as the
canonical Poisson bracket for any Hamiltonian. But a proof appears to be lacking.

\end{document}